\documentclass[%
reprint,
superscriptaddress,
 amsmath,amssymb,
 aps,
 prb,
]{revtex4-1}

\usepackage{graphicx}
\usepackage{dcolumn}
\usepackage{bm}
\usepackage{epstopdf}
\usepackage{subfigure}
\usepackage{underscore}
\usepackage{color}
\usepackage{ulem}
\usepackage{animate}

\begin{document}

\preprint{}

\title{A simple rule for finding Dirac Cones in Bilayered Perovskites}

\author{XueJiao Chen}
\affiliation{
 State Key Laboratory of Luminescence and Applications, Changchun Institute of Optics, Fine Mechanics and Physics, Chinese Academy of Sciences, No.3888 Dongnanhu Road, Changchun 130033, People¡¯s Republic of China
}
\affiliation{
 University of Chinese Academy of Sciences, Beijing 100049, People¡¯s Republic of China
}
\author{Lei Liu}
 \email{liulei@ciomp.ac.cn}
\affiliation{
 State Key Laboratory of Luminescence and Applications, Changchun Institute of Optics, Fine Mechanics and Physics, Chinese Academy of Sciences, No.3888 Dongnanhu Road, Changchun 130033, People¡¯s Republic of China
}
\author{DeZhen Shen}
\email{shendz@ciomp.ac.cn}
\affiliation{
 State Key Laboratory of Luminescence and Applications, Changchun Institute of Optics, Fine Mechanics and Physics, Chinese Academy of Sciences, No.3888 Dongnanhu Road, Changchun 130033, People¡¯s Republic of China
}
\date{\today}

\begin{abstract}
We propose a simple rule for finding Dirac cone electronic states in solids, that is neglecting those lattice atoms inert to the particular electronic bands, and pursuing the two dimensional (2D) graphene-like quasi-atom lattices with s- and p-bindings by considering the equivalent atom groups in the unit cell as quasi-atoms. With CsPbBr$_3$ and Cs$_3$Bi$_2$Br$_9$ bilayers as examples, we demonstrate the effectiveness and generality of this rule with the density functional theory (DFT) calculations. We demonstrate that both bilayers have Dirac cones around the Fermi level and reveal that their corresponding Fermi velocities can reach as high as $\sim$ 0.2$\times$10$^6$m/s. That makes these new 2D layered materials very promising in making new ultra-fast ionic electronic devices.

\end{abstract}

\maketitle

``Dirac Cone'', describing the gapless linear-dispersion of electronic bands, characterizes the superior ballistic massless charge-carrier transport of solids such as in graphene \cite{Wallace1947,Novoselov2005,Zhang2005}, or on the surfaces of topological insulators\cite{Bernevig2006,Hsieh2008,Chen2009}.
Theoretically, the Dirac cones had been predicted in 1947 in graphene, whose honeycomb lattice of $s$-$p$ bonding results in the conical band structure with the linearly dispersive valence and conduction bands touching each other at the Dirac points (K or K$'$) of its hexagonal Brillouin zone\cite{Wallace1947}.
Nevertheless, it was only after its first isolation\cite{Novoselov2004}, graphene has become a source of new sciences, and aroused research upsurges again and again over its novel electronic behaviors\cite{Geim2007,Geim2009,Castro2009,Brumfiel2009,Service2009,NatureNano2010,Kim,Novoselov2012,Service2015,Gibney2018,Cao2018}.
In fact, by energy-band theory and symmetry analysis\cite{Wallace1947}, the normal Dirac cones as presented in graphene will generally appear in those materials, as long as they have the similarly honeycomb bonding style as graphene does.
But, among hundreds of 2D materials examined by now, only graphynes\cite{Malko2018}, silicene and germanene\cite{Cahangirov2009}, ionic boron\cite{Ma2016}, and others\cite{Wang2015} have been identified to be the normal Dirac materials. Nevertheless, unlike graphene, these 2D Dirac materials are made of some sort of artificial lattices where the atoms do not bond together as they do in their stable natural structural polymorphs. In this letter, we propose that following a simple but effective rule more Dirac cone states can be found generally in the 2D materials with stable natural bonding structures.

   \begin{figure}
     \centering
          \includegraphics[width=25em]{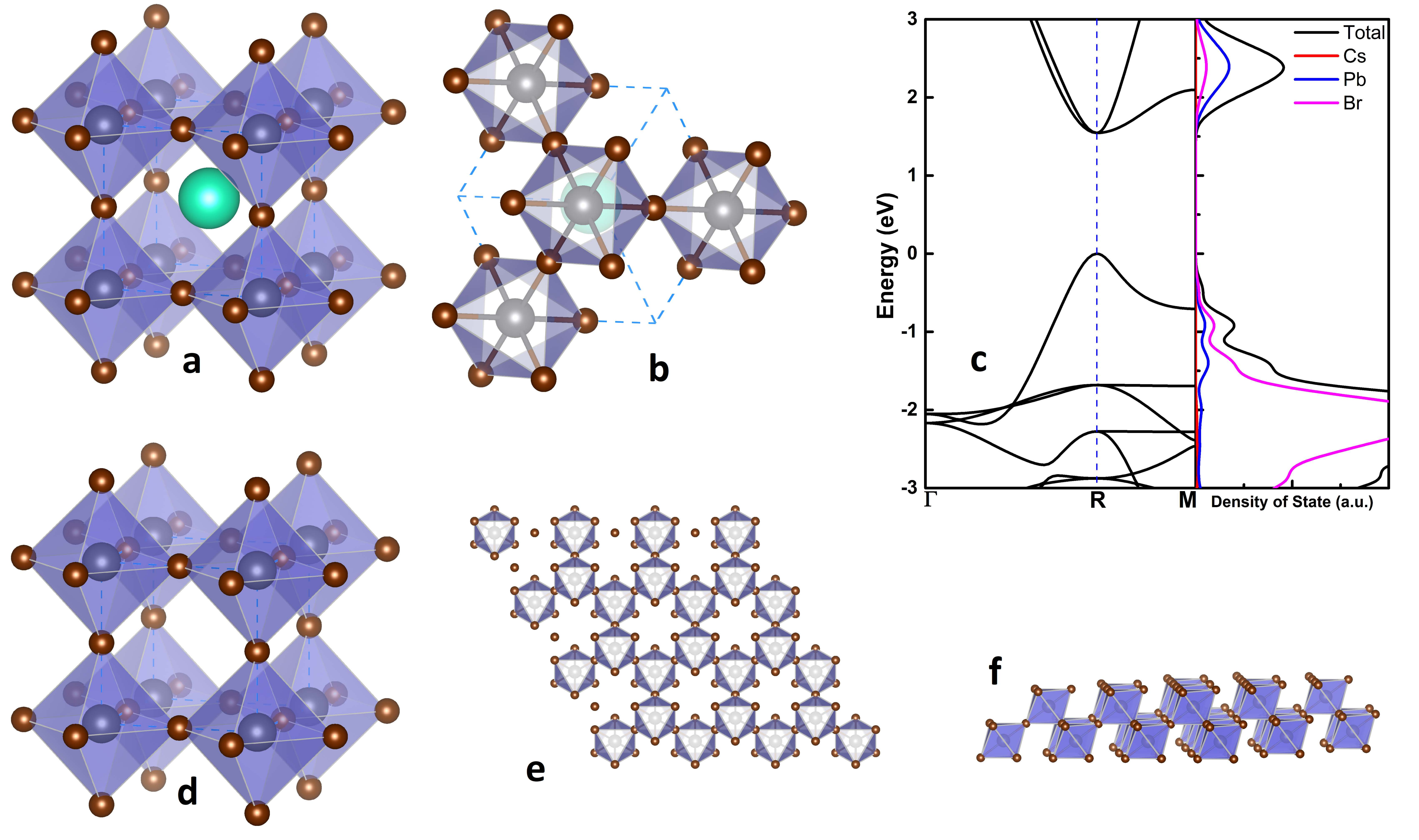}\\
          \caption{\textbf{a} The crystal structure of a cubic ABX$_3$ perovskite, where the green, gray, and brown balls represent the A cations, B cations, and the X anions, respectively; \textbf{b} the crystal structure of a cubic perovskite, viewed from its [111] direction; \textbf{c}  the calculated electronic band structure of bulk CsPbBr$_3$, together with its total and site-decomposed DOS; \textbf{d}  the effective crystal structure of a cubic ABX$_3$ perovskite, effectively seen by its low energy charge carriers; \textbf{e}  the top view of the crystal structure of  a BX$_6$ bilayer sliced out along the (111) plane of a cubic perovskite, which constructs the hexagonal honeycomb lattice with two equivalent  BX$_6$ sublattices; and \textbf{f}  the side view of the crystal structure of  a hexagonal BX$_6$ bilayer, which indeed presents a buckled single quasi-atom (BX$_6$) layer.}\label{fig:fig1}
   \end{figure}

Here, perovskites are selected to demonstrate physics behind the rule we proposed for finding 2D Dirac materials. As shown in Fig. \ref{fig:fig1}a, normally perovskites have the well-known ABX$_3$ lattice structure, where the 6-fold coordinated B cation and its surrounded X anions form the BX$_6$ octahedron, the BX$_6$ octahedra share their X corners forming the 3D skeleton, and the A cations occupy every hole among the 8 BX$_6$ octahedra. For a cubic ABX$_3$ lattice, if viewed from the [111] direction, it presents natively the hexagonal symmetry as shown in Fig. \ref{fig:fig1}b. While the A cations stay isolated from the BX$_6$ skeleton, their electronic states normally do not participate in forming the low-energy bands dispersing near the Fermi level. For a selected halide perovskite (CsPbBr$_3$), Fig. \ref{fig:fig1}c plots its gapped electronic band structure calculated with DFT, together with the total and site-decomposed density of states (DOS). That demonstrates clearly that the electronic orbitals of the A cations stay a few electronvolts away from the valence band maximum (VBM) and the conduction band minimum (CBM) and thus play no role in deciding the low-energy electronic behaviors of a perovskite. Therefore, for those low energy charge carriers, no matter electrons or holes, propagating in a perovskite crystal, the effective lattice they see would be without A cations as shown in Fig. \ref{fig:fig1}d. For such a cubic lattice, if sliced out along the (111) plane, two BX$_6$ layers construct naturally the hexagonal honeycomb lattice with two equivalent BX$_6$ sublattices similar as graphene as shown in Fig. \ref{fig:fig1}e. If considering the BX$_6$ octahedron as a quasi-atom, such a BX$_6$ bilayer transforms into a buckled single quasi-atom layer with structure exactly like silicence\cite{Cahangirov2009} as shown in Fig. \ref{fig:fig1}f.

\begin{figure}
      \centering
      \includegraphics[width=25em]{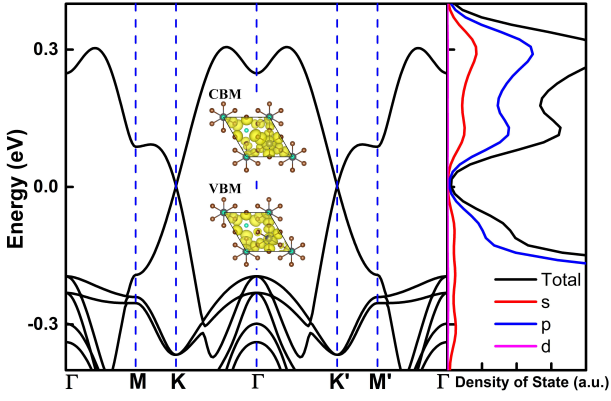}\\
      \caption{The calculated electronic band structure of a hexagonal CsPbBr$_3$ bilayer (the insets plot the isosurfaces of VBM and CBM partial charge densities), together with its orbital-decomposed DOS. }\label{fig:fig2}
\end{figure}

Consequently, a hexagonal perovskite bilayer will present naturally the Dirac-cone electronic states as graphene at its Dirac points of K and K$'$. This is a corollary based on the honeycomb symmetry of BX$_6$ sublattices and the $s$- and $p$-binding characters of BX$_6$ octahedra\cite{Wallace1947}. As a verification example, the hexagonal CsPbBr$_3$ bilayer was selected to calculate its electronic band structures near the Fermi level. Here, the DFT calculations were performed within Perdew-Burke-Ernzerhof (PBE) generalized gradient approximation\cite{pbe} and the projected augmented wave (PAW) method\cite{Bloch1994}, as implemented by the Vienna ab initio simulation package (VASP)\cite{Kresse1994,Kresse1996,Kresse1999}. The cutoff energy for the plane-wave basis set is 300 eV and the Brillouin zone is sampled with the Monkhorst-Pack mesh of 6 $\times$ 6 $\times$ 6 for bulk and 6 $\times$ 6 $\times$ 1 for bilayer CsPbBr$_3$.
The lattice parameter of bulk CsPbBr$_3$ is set to its experimental value a$_0$ = 5.874 {\AA})\cite{Moller1958}. For the CsPbBr$_3$ (111) bilayer, it has the lattice constant of  8.307 \AA \ ($\sqrt2\times 5.874$ \AA) and the thickness of about 6.783 \AA.

As expected, as shown in Fig. \ref{fig:fig2}, its top valence and bottom conduction bands disperse linearly and touch each other at K and K$'$ points. Their orbital-decomposed DOS reveals clearly the $s$- and $p$-orbital nature of both bands around the Dirac points. And that is evidenced further by the insets of Fig. \ref{fig:fig2} which plot the isosurfaces of VBM and CBM partial charge densities. By fitting these two bands at k = K +q to the expression $v_{F}\simeq E(\bf{q})/\hbar|\bf{q}|$\cite{Cahangirov2009}, the Fermi velocity is estimated to be $v{_F}$$\sim$ 0.2$\times$10$^6$m/s, which is one fifth of graphene $v{_F}$. While graphene has the intrinsic carrier mobility of 200000 cm$^2$V$^{-1}$s$^{-1}$\cite{Morozov2008}, it is reasonable to infer that carrier mobility in such a CsPbBr$_3$ bilayer may goes around 40000 cm$^2$V$^{-1}$s$^{-1}$. That makes it rather superior in making ultra-fast electronic devices in comparison with those conventional semiconductors\cite{Yettapu2016}  normally with mobilities less than 10000 cm$^2$V$^{-1}$s$^{-1}$.

\begin{figure}
      \centering
      \includegraphics[width=25em]{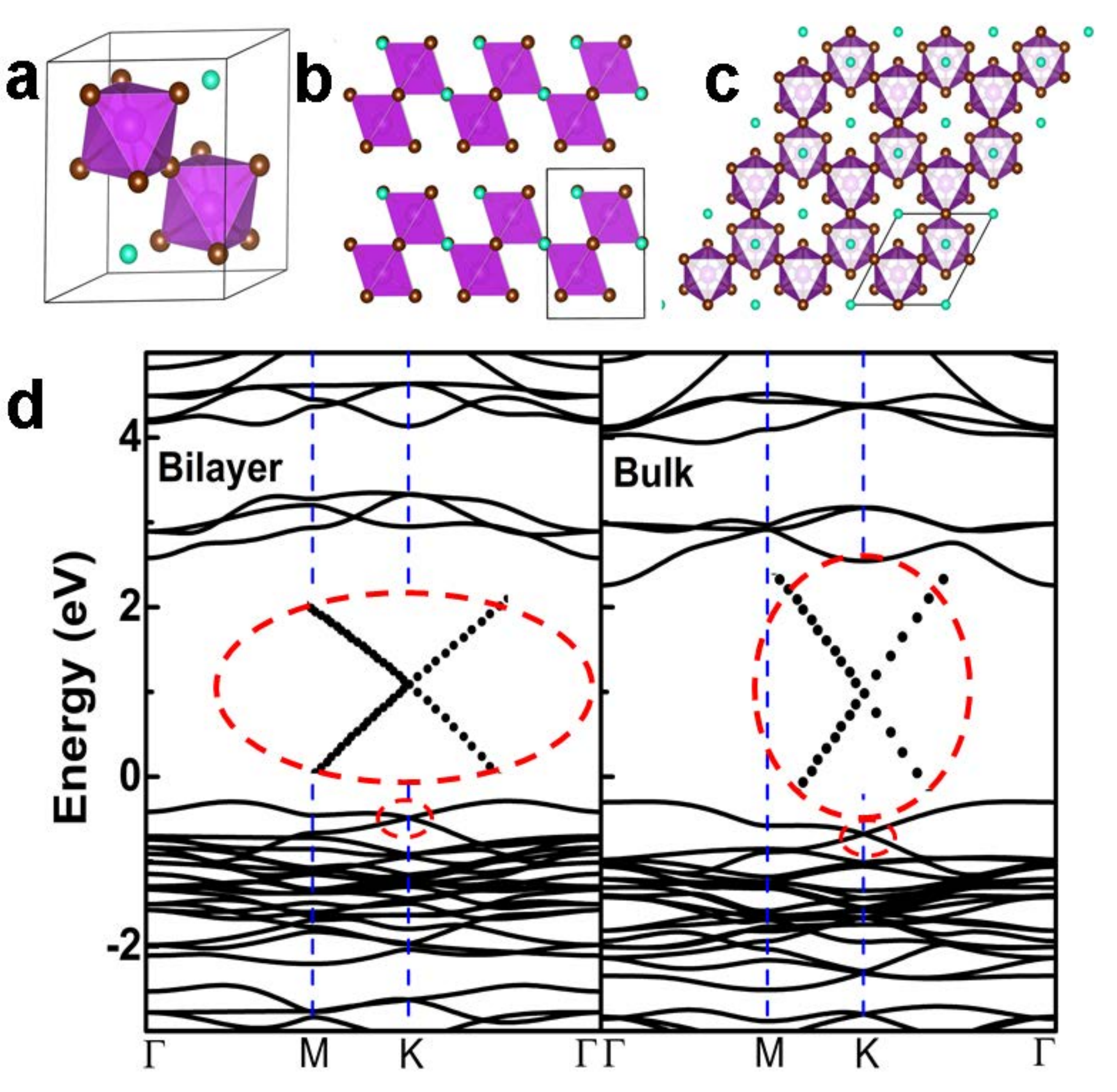}\\
      \caption{The crystal structures of Cs$_3$Bi$_2$Br$_9$ in its primitive cell \textbf{a}, from the side view \textbf{b}, and from the top view along its natural hexagonal axis \textbf{c}; the calculated electronic band structures of bilayer \textbf{d} and  bulk Cs$_3$Bi$_2$Br$_9$ \textbf{e}, where the insets zoom-in their Dirac bands below the Fermi level.}\label{fig:fig3}
\end{figure}

Moreover, the vacancy-order halide perovskites A$_3$B$_2$X$_9$ (A: Cs; B: Fe, Sb, Bi; X: Cl, Br, I) are selected to demonstrate the effectiveness of this rule to find the Dirac electronic states away from the Fermi level. Naturally, the A$_3$B$_2$X$_9$ perovskites are stacked together by those bilayers of linked BX$_6$ octahedra. Figs. \ref{fig:fig3}a and \ref{fig:fig3}b presents the crystal structure of Cs$_3$Bi$_2$Br$_9$, as an example of the vacancy-order halide perovskites. If viewed from their natural hexagonal axis, the A$_3$B$_2$X$_9$ perovskites present the honeycomb lattice of BX$_6$ quasi-atoms, as shown in Fig. \ref{fig:fig3}c. Experimentally, the bulk Cs$_3$Bi$_2$Br$_9$ crystals had been successfully grown by dissolving Bi(OH)$_3$ and Cs$_2$CO$_3$ in a dilute HBr solution, in 1977 by Lazarini et al\cite{Lazarini}.  This layered material presents the trigonal symmetry with space group of P$\bar{3}$m1 and lattice constants of $a = 7.972$ \AA \ and $c = 9.867 $ \AA \cite{Lazarini}.  As a stacking unit, its bilayer is inherently stable as graphene to graphite or single-layer h-BN to bulk h-BN. Based on the similar quasi-atom analysis on the CsPbBr$_3$ bilayer, the Cs$_3$Bi$_2$Br$_9$ bilayer will certainly present the Dirac electronic states as well.

Fig. \ref{fig:fig3}d shows the calculated electronic band structure of the Cs$_3$Bi$_2$Br$_9$ bilayer. And as expected, it presents the crossing Dirac bands as well. However, as Bi contributes one more electron in Cs$_3$Bi$_2$Br$_9$ than Pb does in CsPbBr$_3$, its Fermi level goes above the the Dirac bands for the Cs$_3$Bi$_2$Br$_9$ bilayer. Its Fermi velocity at the crossing point is found to be $v{_F}$$\sim$ 0.2$\times$10$^6$m/s, which equals to the values of the CsPbBr$_3$ bilayer. Furthermore, as the layer interaction is rather weak in bulk Cs$_3$Bi$_2$Br$_9$, its electronic band structure presents the similar Dirac bands below the Fermi lever, as shown in Fig. \ref{fig:fig3}e. It is notable that the electronic band structures of A$_3$B$_2$X$_9$ perovskites have been calculated previously, such as on Cs$_3$Bi$_2$Br$_9$\cite{Bass}, or on Cs$_3$Sb$_2$I$_9$\cite{Singh}. But, in these works, the Dirac-band features of their band structures had been neither noticed or pointed out.

In conclusion, we propose a quasi-atom rule for searching 2D Dirac materials in the conventional solids with natural stable bonding lattices. With this rule, we demonstrate with DFT calculations that hexagonal perovskite ABX$_3$ bilayers, such as the CsPbBr$_3$ (111) bilayer, may have the Dirac band crossing at the Fermi level. Similarly, taking Cs$_3$Bi$_2$Br$_9$ as an example, we demonstrate further the hexagonal vacancy-order halide perovskite A$_3$B$_2$X$_9$ bilayers, together with their bulk phases, will have the Dirac bands below the Fermi level. Both bilayer CsPbBr$_3$ and Cs$_3$Bi$_2$Br$_9$ layers exhibit the Fermi velocity of $\sim$ 0.2$\times$10$^6$m/s. That makes them promising in making ultra-fast electronic devices.

\begin{acknowledgments}
L. L. acknowledges the support from the National Science Fund for Distinguished Young Scholars of China (No. 61525404).
\end{acknowledgments}

\bibliography{Ref}

\pagebreak

\appendix

\titlepage

\setcounter{page}{1}

\begin{center}
{\LARGE Supplemental Information:\\
 A simple rule for finding Dirac Cones in Bilayered Perovskites}
\end{center}
\renewcommand\figurename{SIFIG}
\renewcommand\thefigure{1}
\begin{figure*}[h]
      \centering
      \includegraphics[width=35em]{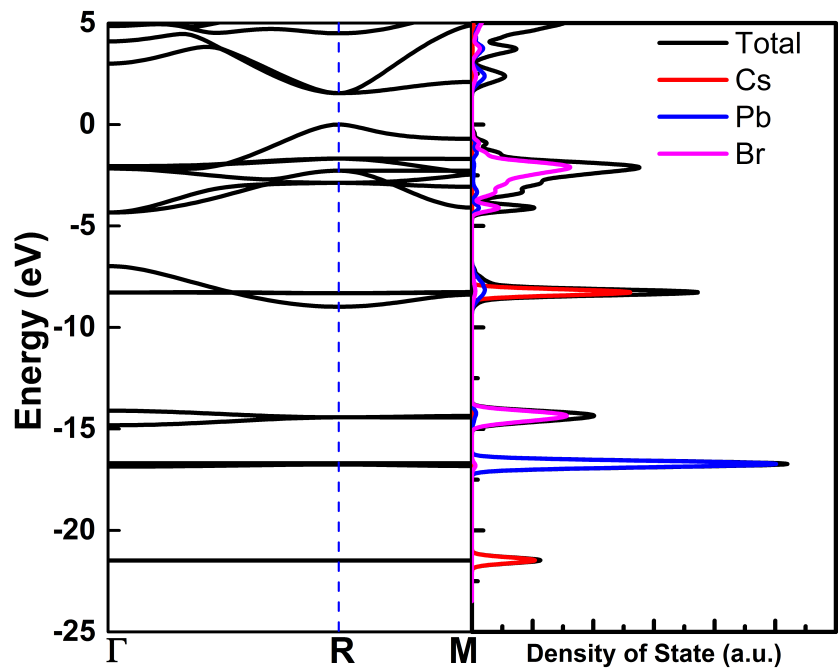}\\
      \caption{The totally calculated electronic band structure of bulk CsPbBr$_3$, together with its orbit-decomposed DOS. }\label{fig:spfig1}
\end{figure*}

\clearpage
\renewcommand\thefigure{2}
\begin{figure*}
      \centering
      \includegraphics[width=35em]{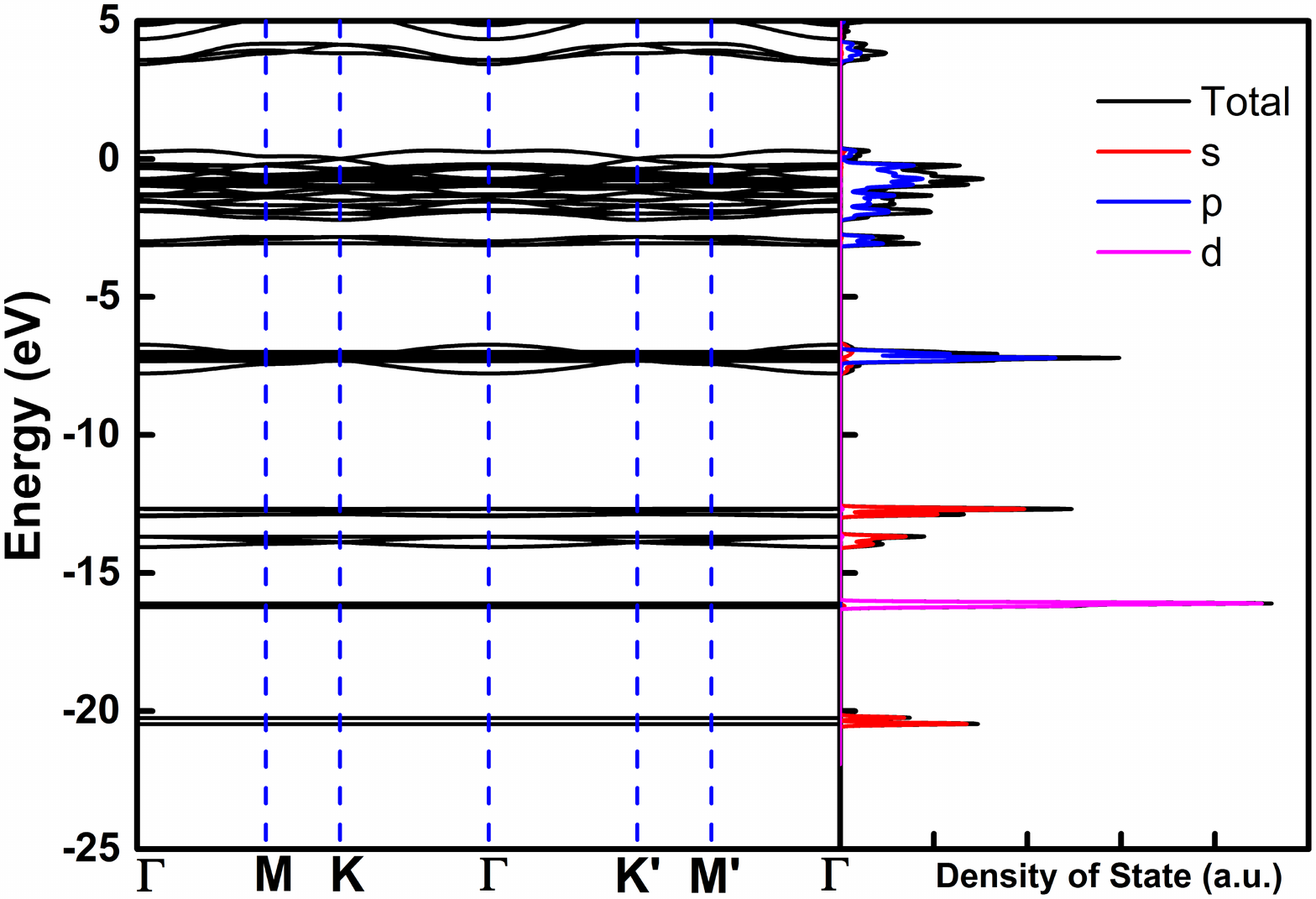}\\
      \caption{The totally calculated electronic band structure of a hexagonal CsPbBr$_3$ bilayer,together with its total and site-decomposed DOS. }\label{fig:spfig2}
\end{figure*}

\end{document}